\input phyzzx.tex
\tolerance=1000
\voffset=-0.0cm
\hoffset=0.7cm
\sequentialequations
\def\rl{\rightline}

\def\t1{{\tilde 1}}

\def\t{\theta}

\REF{\STR}{A. Strominger, JHEP {\bf 0110} (22001) 034; hep-th/0106113.}
\REF{\SSV}{M. Spradlin, A. Strominger and A. Volovich, hep-th/0110007.}
\REF{\VER}{E. Verlinde, hep-th/0008140.}
\REF{\DAN}{U. H. Danielsson, hep-th/0110265.}
\REF{\ADS}{E. Halyo, hep-th/0112093.}
\REF{\KLE}{D. Klemm, hep-th/0106247.}
\REF{\NO}{S. Nojiri and S. D. Odintsov, hep-th/0106191; hep-th/0107134; hep-th/0112152.}
\REF{\DES}{E. Halyo, hep-th/0107169.}
\REF{\INF}{A. Strominger, hep-th/0110087.}
\REF{\SIT}{T. Shiromizu, D. Ita and T. Torii, hep-th/0109057.}
\REF{\CAI}{R. G. Cai, hep-th/0111093; hep-th/0112253.}
\REF{\MED}{A. J. M. Medved, hep-th/0111182; hep-th/0111238; hep-th/0112226.}
\REF{\OGU}{S. Ogushi, hep-th/0111008.}
\REF{\BBM}{V. Balasubramanian, J. de Boer and D. Minic, hep-th/0110108.}
\REF{\PET}{A. C. Petkou and G. Siopsis, hep-th/0111085.}
\REF{\MAN}{A. M. Ghezelbash and R. B. Mann, hep-th/0111217; hep-th/0201004.}
\REF{\CVE}{M. Cvetic, S. Nojiri and S. D. Odintsov, hep-th/0112045.}
\REF{\BMS}{R. Bousso, A. Maloney and A. Strominger, hep-th/0112218.}
\REF{\SPR}{M. Spradlin and A. Volovich, hep-th/0112223.}
\REF{\CAR}{J. L. Cardy, Nucl. Phys. {\bf B270} (1986) 186.}
\REF{\GH}{G. W. Gibbons and S. W. Hawking, Phys. Rev. {\bf D15} (1977) 2738.}
\REF{\BEK}{J. Bekenstein, Lett. Nuov. Cimento {\bf 4} (1972) 737; Phys Rev. {\bf D7} (1973) 2333; Phys. Rev. {\bf D9} (1974) 3292.}
\REF{\HAW}{S. Hawking, Nature {\bf 248} (1974) 30; Comm. Math. Phys. {\bf 43} (1975) 199.}
\REF{\LEN}{L. Susskind, hep-th/9309145.}
\REF{\SBH}{E. Halyo, A. Rajaraman and L. Susskind, Phys. Lett. {\bf B392} (1997) 319, hep-th/9605112.}
\REF{\HRS}{E. Halyo, B. Kol, A. Rajaraman and L. Susskind, Phys. Lett. {\bf B401} (1997) 15, hep-th/9609075.}
\REF{\EDI}{E. Halyo, Int. Journ. Mod. Phys. {\bf A14} (1999) 3831, hep-th/9610068; Mod. Phys. Lett. {\bf A13} (1998), hep-th/9611175.}
\REF{\UNI}{E. Halyo, hep-th/0108167.}
\REF{\HOL}{G. 't Hooft, gr-qc/9310026; L. Susskind, J. Math. Phys. {\bf 36} (1995) 6377, hep-th/9409089.}
\REF{\RAP}{R. Bousso, JHEP {\bf 9907} (1999) 004, hep-th/9905177; JHEP {\bf 9906} (1999) 028, hep-th/9906022.}

\singlespace
\rl{SU-ITP-02-03}
\rl{hep-ph/0201174}
\rl{\today}
\pagenumber=0
\normalspace
\medskip
\bigskip
\titlestyle{\bf{Strings and the Holographic Description of Asymptotically de Sitter Spaces}}
\smallskip
\author{ Edi Halyo{\footnote*{e--mail address: vhalyo@stanford.edu}}}
\smallskip
\centerline {Department of Physics}
\centerline{Stanford University}
\centerline {Stanford, CA 94305}
\centerline{and}
\centerline{California Institute for Physics and Astrophysics}
\centerline{366 Cambridge St.}
\centerline{Palo Alto, CA 94306}
\smallskip
\vskip 2 cm
\titlestyle{\bf ABSTRACT}

Asymptotically de Sitter spaces can be described by Euclidean boundary theories with entropies given by the modified Cardy--Verlinde formula. We show that the Cardy--Verlinde
formula describes a string with a rescaled tension which in fact is a string at the stretched cosmological horizon as seen from the boundary. The temperature of the
boundary theory is the rescaled Hagedorn temperature of the string. Our results
agree with an alternative description of asymptotically de Sitter spaces in terms of strings on the stretched horizon.
The relation between the two descriptions is given by the large gravitational redshift between the boundary and the stretched horizon and a shift in energy.

\singlespace
\vskip 0.5cm
\endpage
\normalspace

\centerline{\bf 1. Introduction}
\medskip

The microscopic origin of the entropy and temperature of (asymptotically) de Sitter spaces is a major puzzle in string theory. Recently, a
holographic description of de Sitter space was given in terms of the de Sitter/CFT correspondence[\STR,\SSV]. In this framework, de Sitter space is described
an Euclidean CFT that lives on its (future or past) boundary. The entropy of space (or the cosmological horizon) can be obtained in the boundary CFT by the
(modified) Cardy--Verlinde formula[\VER,\DAN,\ADS]. This description of de Sitter space suffers from some problems such as the nonunitarity of the boundary CFT, the definition of time
evolution on the Euclidean boundary CFT etc. On the other hand, there seems to be a large amount of supporting evidence in favor of the dS/CFT correspondence[\DAN-\SPR].

The dS/CFT correspondence can be extended to the case of asymptotically de Sitter spaces which are described by nonconformal boundary theories
with a monotonic C--function. Asymptotic de Sitter spaces evolve in (bulk) time into pure de Sitter spaces. On the boundary theory this corresponds
to an RG flow from the IR to the UV which ends at the UV fixed point, i.e. a CFT that describes the de Sitter space. The (modified) Cardy--Verlinde formula
seems to give the correct entropy for the nonconformal boundary theories which are dual to asymptotically de Sitter spaces.

With some simple identifications the (modified) Cardy--Verlinde formula can be put in a form that is very similar to the well--known Cardy formula which holds for two
dimensional, unitary and Lorentzian CFTs[\CAR]. On the other hand, the Cardy--Verlinde formula seems to be valid for Euclidean, nonunitary CFTs (and nonconformal theories)
in more than two dimensions which live on the boundary of (asymptotically) de Sitter spaces. This is a very surprising result which has direct implications for the
validity of the dS/CFT correspondence. The origin of this formula which seems to apply to the above boundary theories is
not understood from first principles. Moreover, its similarity to the Cardy formula seems to imply that the boundary theories are somehow related to two dimensional CFTs.

In this paper we show that a boundary theory for which entropy is given by the (modified) Cardy--Verlinde formula describes a long string with a rescaled tension. We show that
the string actually lives on the stretched horizon of the (asymptotically) de Sitter space.
Seen from the boundary (and taking the rescaling of the boundary radius into account) the tension of the string is rescaled due to the large gravitational
redshift. Similarly, the temperature of the boundary theory corresponds to the rescaled Hagedorn temperature of the string. The entropy of the string (which is the
entropy of the space) is equal to that obtained from the Cardy--Verlinde formula since entropy does not get redshifted. The energy of the boundary theory is related to the
energy of the string at the stretched horizon not only by the gravitational redshift but also by an additive shift.
These results agree with an alternative description of (asymptotically) de Sitter spaces in terms of a long string at the stretched horizon[\DES].
The same string seen from the boundary is described by the boundary theory with the Cardy--Verlinde formula. Each description of (asymptotically) de Sitter spaces has its own
merits. The boundary description allows concrete calculations (such as bulk scattering amplitudes given by boundary correlators); however, it is formulated on the boundary which is not
accessible to an observer who cannot see beyond the cosmological horizon. The string description does not allow detailed calculations but the degrees of freedom which count
the entropy of the space are accessible to an observer by virtue of being inside the horizon.

This paper is organized as follows. In section 2 we give the two alternative microscopic descriptions of asymptotically $dS_3$ spaces and show that the two
descriptions are equivalent. In section 3 we generalize these results to more than three dimensions. Section 4 contains a discussion of our results and our
conclusions.

\medskip
\centerline{\bf 2. Descriptions of Asymptotically $dS_3$ Spaces}
\medskip

In this section we give the two microscopic descriptions of asymptotically $dS_3$ spaces and show that they are equivalent. We first describe the space in terms of
a boundary CFT using the dS/CFT correspondence and show that the entropy of the space is given by the Cardy--Verlinde formula.
We then describe the space in terms of a string with a rescaled tension which lives on the stretched horizon. We find that
the boundary CFT actually describes this string as seen from the boundary with the gravitational redshift of the string energy and tension
taken into account. Thus, the Cardy--Verlinde formula describes the entropy of a string with a rescaled tension living on the stretched horizon. The fact that we
have an effective string is not surprising for $D=3$ since the boundary in this case is two dimensional. On the other hand, there is no a priori reason for expecting
such a relation between the two descriptions.

We now consider asymptotically de Sitter space with $d=3$, i.e. a mass $M$ in $dS_3$ described by the metric[\GH]
$$ds^2=-(1-8G_3 M-{r^2 \over L^2})dt^2+(1-8G_3 M-{r^2 \over L^2})^{-1}dr^2+r^2 d \phi \eqno(1)$$
where $L^{-2}=\Lambda$ and $G_3$ is the three dimensional Newton constant.
This space has only one horizon, namely the cosmological horizon at $r=R<L$ which is given by
$$1-8G_3 M-{R^2 \over L^2}=0 \eqno(2)$$
Thus, $R=L \sqrt{1-8G_3 M}$. Since there is only one horizon, the mass $M$ does not describe a black hole but a stable, pointlike mass which causes a conical defect
in three dimensions.
The temperature of the asymptotically de Sitter space (or the cosmological horizon) is
$$T_{adS}={R \over {2 \pi L^2}}={1 \over {2 \pi L}} \sqrt{1-8G_3 M} \eqno(3)$$
whereas the entropy is given by[\HAW,\BEK]
$$S_{adS}={A \over {4 G_3}}={{\pi L} \over {2G_3}} \sqrt{1-8G_3 M} \eqno(4)$$

The asymptotically $dS_3$ space has two different microscopic descriptions. The first description is in terms of a boundary CFT using the dS/CFT correspondence.
In this case, the CFT lives on the future boundary $I^+$ with a radius of $R$.
The energy of the CFT is (taking into account the redshift by $L/R$)[\VER,\DAN]
$$E_{CFT}={c \over {48 \pi}} {V \over L^2}(1-{L^2 \over R^2}) \eqno(5)$$
where $V=2 \pi R$ is the volume of the one dimensional boundary and $c=3L/G_3$ is the central charge. Above, the first term is the extensive and positive
contribution of the CFT gas ($E_E$)
whereas the second term gives the sub--extensive and negative Casimir energy due to the finite volume ($E_C$).
The central charge $c$ and the level $L_0$ of the CFT are defined in terms of the Casimir
energy $E_C$ and the total energy $E_{CFT}$ respectively
$$c=24 R E_C \qquad L_0=R E_{CFT} \eqno(6)$$
The entropy is given by the modified Cardy--Verlinde formula[\VER,\DAN]
$$S_{CFT}=4 \pi R \sqrt{E_E |E_C|} \eqno(7)$$
which reproduces the Bekenstein--Hawking result in eq. (4). With the identifications in eq. (6) $S_{CFT}$ can be put in a form that is very similar to the Cardy formula.
The temperature of the CFT is obtained from
$$T_{CFT}=\left( \partial E_{CFT} \over \partial S_{CFT} \right)={1 \over {2 \pi L}} \eqno(8)$$
which is related to the temperature of the bulk by the redshift $T_{CFT}=(L/R)T_{adS}$.
Therefore, the asymptotically $dS_3$ space is described by a thermal state of a (two dimensional) boundary CFT with central charge $c$.

Using $E_C=(L^2/R^2)E_E$ we can write the entropy of the CFT as
$$S_{CFT}=4 \pi L E_E \eqno(9)$$
We see that the entropy is proportional to the energy which is a characteristic of string--like behavior.
Thus, eq. (9)  can be seen as the entropy of a string with ($c=6$ and) tension $T_{str}=1/2 \pi L^2$ and energy $2 E_E$. This by itself is not surprising since the boundary
in this case is two dimensional. However, below we show that this CFT describes a string on the stretched horizon seen from the boundary which corresponds to an alternative
description of asymptotically $dS_3$ space.

The second description of the asymptotically $dS_3$ space is given by a long string with a rescaled tension which lives on the stretched horizon. This type of
description has been useful in explaining the entropy of all black objects with Rindler--like near horizon geometries[\LEN-\UNI] and de Sitter space in all dimensions[\DES].
We take the near horizon limit by $r=R-y$ where $y<<R$. Then we find
$$1-8G_3 M-{r^2 \over L^2} \to {{2 Ry} \over L^2} \eqno(10)$$
The proper distance to the horizon becomes
$$\rho=\sqrt{2y \over R}L \eqno(11)$$
Then, the near horizon metric describes Rindler space
$$ds^2=-({R^2 \over L^4}) \rho^2 dt^2+d\rho^2+R^2 d\phi \eqno(12)$$
We find the dimensionless Rindler time
$$\tau_R={R \over L^2} t \eqno(13)$$
Using
$$[E_R,\tau_R]=1=[E_R,t]{R \over L^2} \eqno(14)$$
we find the Rindler energy
$$E_R={L \over {4G_3}} \sqrt{1-8G_3 M} \eqno(15)$$
We see that the Rindler energy is the entropy of the cosmological horizon; $S_{adS}=2 \pi E_R$. A microscopic description of this result can be given in terms of a long
string (with tension $T_{str}=1/2 \pi \ell_s^2$ and $c=6$) which lives on the stretched horizon at $r=R-\ell_s$. Then the energy of such a string would be $E_{sh} =E_R/\ell_s$.
Therefore, this string is in a state with level number $n= E_R^2$ which means the entropy of the string is the entropy of the horizon $S_{str}= 2 \pi \sqrt{c n/6} =
2 \pi E_R=S_{adS}$. Since $T_R=1/2 \pi$, the temperature on the stretched horizon is $T_{sh}= 1/2 \pi \ell_s$, i.e the Hagedorn temperature of the string. However,
a cosmological observer (with time $t$) will
see a rescaled energy, tension and temperature due to the large gravitational redshift near the horizon. Using eq. (13) we find that the rescaled energy is
$E_{cos}= (1-8G_3 M)/4 G_3$ whereas the rescaled temperature is $T_{cos}= \sqrt{1-8G_3 M}/2 \pi L=T_{adS}$. Seen by the cosmological observer, the string on the stretched
horizon has a rescaled tension $T_{str}= (1-8G_3 M)/2 \pi L^2$. The entropy being a scalar does not redshift and therefore $S_{str}=S_{adS}$.

We now argue that the above two descriptions are equivalent by showing that the effective string that the boundary CFT describes is the string at the stretched horizon.
Consider the description of the long string as seen by an observer at the boundary. It is easy to see that (taking into account the
coordinate rescaling of the boundary to get a radius of $R$) the redshift between the stretched horizon and the boundary is $\ell_s/L$. Then a string with a scale $\ell_s$
will be seen as a string with scale $L$ from the boundary.
Thus, on the boundary the string tension is renormalized to $T_{str}=1/2 \pi L^2$ due to the large gravitational redshift. The temperature of the boundary CFT is given
by the redshifted Hagedorn temperature of the string, $T_{CFT} =1/2 \pi L$. The entropy of the boundary theory (i.e.
the entropy of the cosmological horizon) is the entropy of the string which does not get redshifted.
We note that for the energy there is not only a (multiplicative) redshift but also an additive shift between the boundary and the stretched
horizon. This can be easily seen for the pure de Sitter case (with $M=0$) which gives $E_{CFT}=0$ but $E_{sh}= L/4 G_3 \ell_s$. We find that the shift in energy
(on the boundary) is precisely $2 E_C$. Note that the energy shift remains $2E_C$ for the asymptotically $dS_3$ space with $M>0$ even though $E_C$ changes due to its
$R$ dependence.
This means a string on the stretched horizon with $E_{sh}= E_R/\ell_s$ corresponds on the boundary to a state with
$2 E_E=(\ell_s/L)E_{sh}$ exactly as given by eq. (5). This shift in energy is very similar to the shift between energies as measured in cosmological and static
coordinates. We do not have an explanation for the fact that the two descriptions of (asymptotically) de Sitter spaces relate to two different energies.

We found that the boundary CFT description of asymptotically $dS_3$ space is another way of describing a long string on the stretched horizon as seen from the
boundary. Therefore the two microscopic descriptions of asymptotically $dS_3$ spaces seem to be equivalent. In the above case, the stringy behavior is not very surprising
since the boundary of asymptotically $dS_3$ spaces are two dimensional, i.e. string--like. However, the fact that the two dimensional boundary theory describes a string near the
horizon is highly nontrivial.

\medskip
\centerline{\bf 3. Generalization to Higher Dimensions}
\medskip

The results of the previous section can be easily generalized to higher ($D>3$) dimensions. For $D>3$ the metric of asymptotically de Sitter space with a black hole
of mass $M$ is[\GH]
$$ds^2=-(1-{r^2 \over L^2}-{\mu^d \over r^d})dt^2+(1-{r^2 \over L^2}-{\mu^d \over r^d})^{-1}dr^2+r^2 d \Omega_{d+1}^2 \eqno(16)$$
where $\mu=G_D M$ and $D=d+3$. Now there are two horizons and their positions are given by the positive roots of
$$1-{r^2 \over L^2}-{\mu^d \over r^d}=0 \eqno(17)$$
The larger root which we denote by $R$ gives the position of the cosmological horizon whereas the smaller one gives the position of the black hole horizon.
The temperature and entropy of the cosmological horizon are given by
$$T_{adS}={R \over {4 \pi L^2}}((d+2)-d{L^2 \over R^2}) \eqno(18)$$
and
$$S_{adS}={A \over {4 G_D}}={R^{d+1} \over 4 G_D} \eqno(19)$$

As for the $D=3$ case, the asymptotically $dS_D$ space with $D>3$ can be described by a boundary field theory using the dS/CFT correspondence.
The energy of the boundary theory is given by
$$E_{BND}={{c (d+1)}\over {48 \pi}} {V \over L^{d+2}}(1-{L^2 \over R^2}) \eqno(20)$$
where the first term is the positive, extensive contribution
of the CFT gas ($E_E$) and the second term is the negative, sub--extensive Casimir energy ($E_C$).
In $D$ dimensions we have $c=3L^{d+1}/G_D$. This boundary
theory was shown to be nonconformal with a monotonic C--function defined by
$$48  \pi R E_C \leftrightarrow {(d-2)C} \qquad 2 \pi R E_{BND} \leftrightarrow (d-2){\tilde L_0} \eqno(21)$$
The temperature of the boundary theory is given by $T_{BND}=(L/R)T_{CFT}$.
The entropy of the boundary theory is given by the Cardy--Verlinde formula[\VER,\DAN]
$$S_{BND}={{4 \pi R} \over {(d+1)}} \sqrt{E_E |E_C|} \eqno(22)$$
Again, using eq. (21) $S_{BND}$ can be put in a form very similar to the Cardy formula even though we do not have a conformal boundary theory.
The asymptotically de Sitter space (with a black hole) is not stable due to the Hawking radiation from the black hole which is at a higher temperature that the surrounding
space. It has been argued that time evolution of the space corresponds on the boundary theory to an RG flow from the IR to the UV. After the black hole evaporates and all radiation
leaves the cosmological horizon, the space
becomes pure de Sitter. On the boundary this corresponds to the RG flow reaching the UV fixed point which is a CFT that describes the de Sitter space.
As before, using $E_C=(L^2/R^2)E_E$ we find
$$S_{BND}={{4 \pi} \over {(d+1)}} L E_E \eqno(23)$$
Thus, we find a string--like behavior even for $D>3$ where the boundary has a dimension higher than two. Eq. (23) seems to describe a string with tension
$T =P^2/2 \pi L^2$ and energy $2 P E_E/(d+1)$ where $P$ is a constant. For $D=3$ we found in the previous section that $P=1$; however this may not be the case for $D>3$.
In fact, we will see below that $P$ is a function of $R$, $L$ and $d$ which is determined by the gravitational redshift factor from the cosmological stretched horizon to the boundary.
This is very intriguing and may be the reason behind the seemingly two dimensional nature of Cardy--Verlinde formula for boundary theories in any dimension.

We now describe the asymptotically de Sitter space by a long string with a rescaled tension which lives on the cosmological stretched horizon. This can be considered as a
generalization of ref. [\DES] which described the entropy of de Sitter space by the same methods. In the near horizon region,
$r=R-y$ with $y<<R$. Then the near horizon metric becomes
$$ds^2=-2\left({{(d+2)R^2-dL^2} \over {RL^2}} \right) ydt^2+{1 \over {2y}} \left({{(d+2)R^2-dL^2} \over {RL^2}}\right)^{-1}dr^2+r^2d \Omega_{d+1}^2 \eqno(24)$$
The proper distance to the horizon is
$$\rho=\sqrt {2y} \left({{dL^2-(d+2)R^2} \over {RL^2}} \right)^{-1/2} \eqno(25)$$
The near horizon metric describes Rindler space
$$ds^2=-{1 \over 4} \left({{(d+2)R^2-dL^2} \over {RL^2}} \right)^2 \rho^2 dt^2+d\rho^2+R^2d \Omega_{d+1}^2 \eqno(26)$$
The dimensionless Rindler time is given by
$$\tau_R={1 \over 2} \left({{(d+2)R^2-dL^2} \over {RL^2}} \right) t \eqno(27)$$
Note that this reduces to eq. (13) for $D=3, d=0$ as expected. Using eqs. (14) and (27) we find the Rindler energy
$$E_R = {R^{d+1} \over {4 G_D}} \eqno(28)$$
Again we observe that $E_R$ is proportional to the entropy of the cosmological horizon; $S_{adS}=\Omega_{d+1} (d+1) E_R$ where $\Omega_{d+1}$ is the volume of
the unit $S^{d+1}$. As before, this
result can be microscopically explained in terms of a long string (with tension $T=1/2 \pi \ell_s^2$ and $c=6$)
situated at the stretched cosmological horizon. The energy of the string is $E_{sh}=E_R/\ell_s$ and its entropy is equal to that of the cosmological horizon. A cosmological
observer with time $t$ sees
the string tension rescaled due to the gravitational redshift to $T_{str}=((d+2)R^2-dL^2)^2/8 \pi R^2L^4$. The temperature of the space is given by the rescaled Hagedorn
temperature of the string, $T_{adS}=((d+2)R^2-dL^2)/4 \pi RL^2$ which agrees with eq. (18).

We now argue, as in the $D=3$ case, that the two descriptions of asymptotically $dS_D$ space are equivalent.
Seen from the boundary, the energy and temperature of the string on the cosmological stretched horizon are rescaled by the redshift factor $\ell_s ((d+2)R^2-dL^2)/R^2 L$.
The entropy does not get rescaled so we have $S_{BND}=S_{str}=S_{adS}$. We see that the boundary theory describes a string with a tension
$((d+2)R^2-dL^2)^2/8 \pi R^4 L^2$. The temperature of the boundary theory is the rescaled Hagedorn temperature of the string, i.e. $T_{BND}=((d+2)R^2-dL^2)/ 4 \pi R^2L$
which agrees with $T_{adS}$ in eq. (18) up to the redshift factor.
The energy of the boundary theory becomes $E=E_R (d+2)R^2-dL^2)/2 R^2 L$ which is related to $E_{BND}$ in eq. (21) by a shift which depends on $R$.

We found that the boundary theory in which the entropy is given by the Cardy--Verlinde formula actually describes a string with a rescaled tension. This is very
surprising since the boundary for $D>3$ is not two dimensional. It seems that this relation between the boundary theory and the string on the stretched horizon is the origin
of the apparent two dimensional nature of the Cardy--Verlinde formula for $D$ dimensional boundary theories.

\medskip
\centerline{\bf 5. Conclusions and Discussion}
\medskip

In this paper we showed that the boundary theories which are dual to asymptotically de Sitter spaces describe strings with rescaled tensions. This can be easily seen
from the Cardy--Verlinde formula which shows that the entropy is proportional to the energy. The string that the boundary theory describes lives
at the cosmological stretched horizon and its tension is rescaled due to the large gravitational redshift between the boundary and the horizon. The temperature of the
boundary theory is the rescaled Hagedorn temperature of the string. The energy of the boundary theory and that of the string are related by
the gravitational redshift and an additive shift. The reason for the appearance of two different energies in these descriptions is not clear to us. This picture
agrees with an alternative description of asymptotically de Sitter spaces in terms of strings with rescaled tension at the stretched horizon. The two descriptions
are related by the gravitational redshift between the boundary and the stretched horizon. The boundary description of the space allows for computation of scattering
amplitudes etc. but is not accessible to any observer since it takes place beyond the cosmological horizon. The string decription is not suitable for explicit
calculations but describes degrees of freedom that can be observed since they are inside the cosmological horizon. In this sense, the two descriptions seem to
complement each other.

An important feature of asymptotically de Sitter spaces is their instability due to Hawking radiation from the black hole. On the boundary theory this is described
by the RG group flow from the IR to the UV. How is this represented in the description with the string at the stretched horizon? In this description, each horizon is described
by a string that lives at the respective stretched horizon. The string at the stretched black hole
horizon has a higher temperature than the one at the stretched cosmological horizon (as seen by a cosmological observer) and therefore it emits radiation (small closed
strings) which are absorbed by the string at the stretched cosmological horizon. Due to this absorbtion of energy the string at the stetched cosmological horizon gets longer
which effectively increases the radius of the cosmological horizon.

In the description with the strings there is a symmetry between the two horizons; they are both described by strings in the same manner. On the other hand,
it is known that in the framework of the dS/CFT correspondence, a description of the black hole horizon in terms of a boundary theory is lacking. Perhaps, the above
relation between the two descriptions of asymptotically de Sitter space is an indication that the black hole horizon is described by a(nother) boundary theory on $I^-$.

It would be interesting to see if the above equivalence betwen the two descriptions can be generalized to flat space.
Black objects in flat space--time can be described by strings with rescaled tensions living on stretched horizons. A naive extrapolation of our results show that, in the flat
space--time case, the boundary goes to infinity and the central charge of the boundary theory diverges. It seems difficult to make sense out of such a theory. On the other hand,
one can speculate that the holographic theory lives on the horizon of the black object with a central charge fixed by the black hole (and not the cosmological) radius.

Our results indicate that there is a clear relation between strings that live on stretched horizons and the holographic field theories that are dual to a given space--time[\HOL,\RAP].
In this paper we showed this for asymptotically de Sitter spaces and their holographic duals. In order to continue along these lines, one needs to better understand the field
theories in question. This is problematic for asymptotically de Sitter spaces since the field theories on the boundary are Euclidean and not unitary. On the other hand, it seems
that all our results hold equally well to an AdS black hole in the context of of AdS/CFT correspondence.
It would be worthwile to examine the role of strings on stretched horizons in AdS space and their relation to the boundary theories which are better understood.

\bigskip
\centerline{\bf Acknowledgements}

I would like to thank the referee of hep-th/0107169 for pointing out the possibility of energy shifts between the two descriptions of de Sitter spaces.

\vfill

\refout

\end
\bye